\documentclass[sigconf,nonacm]{acmart}

\usepackage{algorithm} 
\usepackage{algpseudocode}

\AtBeginDocument{%
  }

\setcopyright{acmlicensed}
\copyrightyear{2018}
\acmYear{2018}
\acmDOI{XXXXXXX.XXXXXXX}

\acmConference[Conference acronym 'XX]{Make sure to enter the correct
  conference title from your rights confirmation emai}{June 03--05,
  2018}{Woodstock, NY}
\acmISBN{978-1-4503-XXXX-X/18/06}

\begin{document}

\title{NV-Retriever: Improving text embedding models with effective hard-negative mining}

\author{Gabriel de Souza P. Moreira*}
\affiliation{%
  \institution{NVIDIA}
  \city{S\~ao Paulo}
  \country{Brazil}}
\email{gmoreira@nvidia.com}

\author{Radek Osmulski*}
\affiliation{%
  \institution{NVIDIA}
  \city{Brisbane}
  \country{Australia}
}
\email{rosmulski@nvidia.com} 

\author{Mengyao Xu*}
\affiliation{%
 \institution{NVIDIA}
 \city{Santa Clara}
 \country{USA}}
\email{mengyaox@nvidia.com}  

\author{Ronay Ak*}
\affiliation{%
  \institution{NVIDIA}
  \city{Sarasota}
  \country{USA}}
\email{ronaya@nvidia.com}  

\author{Benedikt Schifferer}
\authornote{All authors contributed equally to this research.}
\affiliation{%
  \institution{NVIDIA}
  \city{Berlin}
  \country{Germany}
}
\email{bschifferer@nvidia.com}

\author{Even Oldridge*}
\affiliation{%
  \institution{NVIDIA}
  \city{Vancouver}
  \country{Canada}}
\email{eoldridge@nvidia.com}

\renewcommand{\shortauthors}{Moreira et al.}

\begin{abstract}
  Text embedding models have been popular for information retrieval applications such as semantic search and Question-Answering systems based on Retrieval-Augmented Generation (RAG). 
  Those models are typically Transformer models that are fine-tuned with contrastive learning objectives.
One of the challenging aspects of fine-tuning embedding models is the selection of high quality hard-negative passages for contrastive learning. 
  In this paper we introduce a family of positive-aware mining methods that use the positive relevance score as an anchor for effective false negative removal, leading to faster training and more accurate retrieval models. We provide an ablation study on hard-negative mining methods over their  configurations, exploring different teacher and base models. 
  We further demonstrate the efficacy of our proposed mining methods at scale with the NV-Retriever-v1 model, which scores 60.9 on MTEB Retrieval (BEIR) benchmark and placed 1st when it was published to the MTEB Retrieval on July, 2024.

\end{abstract}



\maketitle

\section{Introduction}
Text retrieval is critical for a range of information retrieval applications such as search, question answering, semantic textual similarity, and item recommendation. It is also vital to the field of Retrieval Augmented Generation (RAG)\cite{lewis2020retrieval,ram2023context}, which enables Large Language Models (LLM) to  access external context without modifying model parameters. 

Dense embedding models are a key component of text retrieval, semantically representing queries and passages (pieces of content) with low token overlap and generalize for out-of-domain corpus. Retrieval systems that index passages into embeddings can efficiently retrieve relevant passages for a query using Maximum Inner Product Search (MIPS) \cite{lewis2020retrieval}.

There has been a growing interest in text embedding models from both academia and industry leading to the recent release of many models including E5\cite{wang2022text}, GTE\cite{li2023towards}, and Jina\cite{gunther2023jina}. To consistently compare the accuracy of available text embedding models the MS-MARCO\cite{bajaj2016ms}, BEIR\cite{thakur2021beir} and MTEB\cite{muennighoff2022mteb} benchmarks on public (HuggingFace\cite{wolf2019huggingface}) leaderboards provide an important focal point of comparison for different embedding based tasks.

Embedding models are trained with Contrastive Learning (CL)\cite{chen2020simple} to maximize the similarity between the embeddings of a query and the relevant passages  (positives) while minimizing the similarity with embeddings of passages not relevant to the query (negatives)\cite{karpukhin2020dense}. 

When preparing data for training embedding models, \textit{hard-negative mining} is typically used to select negative passages for queries. It leverages a teacher retrieval model to find passages somewhat relevant to the query, making it harder for the contrastive loss to differentiate positives from negatives, resulting in more efficient and effective fine-tuning of the embedding model.

Despite its importance for embedding model fine-tuning, the \textit{hard-negative mining} methods remain underexplored or poorly detailed, particularly in papers \cite{wang2023improving,LinqAIResearch2024,li2023towards, li2023towards, nussbaum2024nomic} introducing models topping the MTEB leaderboard, as these studies primarily focus on model architectures, fine-tuning approaches and data blends for training. 

The main contributions of this research are threefold:
\begin{itemize}
    \item \textbf{Positive-aware hard-negative mining methods}. We introduce a family of mining methods that are capable of improving contrastive learning, removing potential false negatives and increasing the accuracy of text embedding models, as described in Section~\ref{sec:mining_methods};
    
    \item \textbf{Investigation on hard-negative mining best practices}. We present in Section~\ref{sec:methodology} our methodology and research questions, and in Section~\ref{sec:exp_models} compare different hard-negative mining methods over their configurations, with different teacher and base models, demonstrating how sensitive retrieval models are to hard-negative mining choices;
    
    \item \textbf{Scaling the mining methods to achieve state-of-the-art text retrieval: \textit{NV-Retriever-v1}}. We demonstrate in Section~\ref{sec:nv_retriever_ablation} the effectiveness of our proposed mining methods at scale in the training of \textit{NV-Retriever-v1}, a state-of-the-art embedding model which achieved 1st place on MTEB Retrieval leaderboard\footnote{https://huggingface.co/spaces/mteb/leaderboard} when published on MTEB on July 11th, 2024. We describe \textit{NV-Retriever-v1} train blends, hyperparameters, architecture, and particularly how our positive-aware hard-negative mining methods were crucial to its leading retrieval accuracy.
    
\end{itemize}

\section{Background}

In this section we discuss related work on text embedding models and hard-negative mining.

\subsection{Text embedding models}

Text embedding models represent variable-length text as a fixed dimension vector that can be used for downstream tasks.

A seminal work on embedding sentences was Sentence-BERT \cite{reimers2019sentence}, which proposed a modification to BERT network to represent pairs of related short texts in the same embedding space by using siamese networks (query,positive passages) or triplet networks (query,positive,negative passages). They also explored different options of objective functions and embedding pooling.

Contrastive learning was popularized by SimCLR\cite{chen2020simple} superior performance than classification-based losses\cite{reimers2019sentence} for embeddings. The (DPR)\cite{karpukhin2020dense} proposed a bi-encoder architecture where separate BERT encoders (with no shared weights) represented query and passage, whose output embeddings are used for CL. 

The E5 \cite{wang2022text} family of models leveraged two-stage training of embedding models, pre-training with unsupervised pairs of text (e.g. neighboring text spans, title-abstract) and fine-tuning with supervised  data (e.g. question-answer, text-summary, search-relevant passages). E5 models are available in different sizes, depending on the base model: MiniLM\cite{wang2020minilmv2} and BERT\cite{devlin2018bert}. E5 was the first dense retrieval model to beat BM25\cite{robertson2009probabilistic} sparse  model baseline on BEIR\cite{thakur2021beir} without using labeled data (with its unsupervised version). On second stage, they fine-tune E5 unsupervised model with labeled data from MS-MARCO dataset \cite{bajaj2016ms}, Natural Questions (NQ) and NLI for superior performance.

E5-Mistral \cite{wang2023improving} embedding model proposed using a decoder model instead of encoder model (like BERT) as a base model. They choose Mistral-7B\cite{jiang2023mistral}, which is already extensively pre-trained on web-scale data. It was fine-tuned in a single round with synthetic data generated by LLMs for different tasks and languages and with a small amount of labeled data.

The BEIR benchmark\cite{thakur2021beir} has become the standard evaluation for zero-shot text retrieval with 18 retrieval datasets. Later MTEB \cite{muennighoff2022mteb} was introduced as a more comprehensive text embeddings benchmark, with 56 datasets distributed over 7 tasks for the English subset: retrieval, reranking, classification, clustering, pair classification, summarization, and semantic textual similarity. The MTEB retrieval task is composed by 15 selected datasets from BEIR. We evaluate on MTEB Retrieval in Section~\ref{sec:nv_retriever_ablation}.

\subsection{Hard-negative mining for fine-tuning embedding models}
\label{sec:hard_neg_mining_background}

Contrastive Learning (CL) requires triples of query, positive passage and negative passages. Negative passages can be either labeled by humans or, more commonly be sampled from the corpus.

A basic approach for selecting negative passages is using the positive passages from other examples (queries) in the batch (\textit{in-batch negatives}) \cite{chen2020simple,karpukhin2020dense}. This is computationally efficient because the embeddings for those passages were already generated by the model forward pass, although, the batch size limits the number of negatives. Some proposals to increase the number of negatives are keeping a memory bank with embeddings from past batches  \cite{xiong2020answering,wang2022text} or combining batches from different GPUs \cite{qu2020rocketqa} (\textit{cross-batch}).

The \textit{in-batch} negatives are random with respect to the query and easy to discriminate. Thus, they are very uninformative for CL, as their loss / gradients are low and contribute little to model convergence \cite{xiong2020approximate}. On the other hand, using challenging (\textit{hard}) negatives can lift the upper bound of gradient norm, reduce the variance of stochastic gradient estimation, and lead to faster learning \cite{xiong2020approximate}. 

Hard-negatives can be mined from the corpus of passages by retrieving examples similar to the query that are not labeled  positive\cite{gillick2019learning,wu2019scalable,karpukhin2020dense,xiong2020approximate}. Sparse or dense (embedding) retrieval models can be used for mining. DPR \cite{karpukhin2020dense} uses one or two \textit{hard} negatives mined with BM25\cite{robertson2009probabilistic} in addition to the in-batch negatives that help them to finetune models with better accuracy.

Some methods have been proposed for incremental hard-negative mining during training. ANCE\cite{xiong2020approximate} asynchronously refreshes the ANN index with the updated passage embeddings and re-mines the hard-negatives for each question. 
NGAME\cite{dahiya2023ngame}  clusters queries and positive embeddings and uses those clusters to prepare \textit{negative mining-aware mini-batches}. The  batches contain data points close to each other in the embedding space, so that hard-negatives can be found among the in-batch samples, whose embeddings are efficiently computed. Incremental mining methods such as ANCE and NGAME are in general complex to implement and costly to compute for large corpus, which is why for most of the top models in MTEB \cite{LinqAIResearch2024, li2023towards, nussbaum2024nomic,wang2023improving} the negatives are mined upfront (before training) using a pre-trained model.

\subsubsection{False negatives}

In \cite{qu2020rocketqa} they found that naive mining of hard-negatives may select many false negatives. Their experiment on mining negatives from the MS-Marco dataset found that about 70\% of passages most similar to the queries should be actually labeled as positive. They propose to \textit{denoise} hard-negatives, i.e., ignore the potential false negatives by filtering out the ones with high relevance score to the query.

Some works try to denoise hard-negatives retrieved from embedding models with cross-encoder models, like in RocketQA \cite{qu2020rocketqa, ren2021rocketqav2}, or with more powerful decoder LLMs\cite{lee2024gecko}. These approaches might be costly to run on large train sets depending on the model size, as they require inference for every (query, negative) pair.

While most top performing models on MTEB like \textit{e5-mistral-7b-instruct} \cite{wang2023improving}, \textit{Linq-Embed-Mistral}
\cite{LinqAIResearch2024}, \textit{NV-Embed-v1} \cite{lee2024nv}, \textit{gte-large-en-v1.5} \cite{li2023towards}, \textit{nomic-embed-text-v1} \cite{nussbaum2024nomic} leverage hard-negative mining in their fine-tuning, they do not explore or describe in detail their methodology to decide which model and method to use for mining. In \textit{snowflake-arctic-embed-l} \cite{merrick2024arctic}, the authors include an ablation study of three maximum negative score thresholds (0.4, 0.5 and 0.8) for hard-negative mining. The \textit{SFR-Embedding-Mistral} \cite{SFRAIResearch2024} blog provides an ablation on sampling hard-negatives from three different ranges of top-k candidates (0-100, 30-100, 50-100) and reports that the range 30-100 helps eliminate false negatives and improve model performance. They also compare different teacher models for mining (BM25, BGE-base \cite{bge_embedding}, E5-Mistral\cite{wang2023improving} and their own SFR-Embedding-Mistral) and show that more powerful models can yield more effective hard-negatives, which we also observe in our ablation presented in Section~\ref{sec:exp_models}.


\section{Methodology}
In this section, we introduce our main contribution - the positive-aware hard-negative mining methods - the research questions, and experiments setup.

\subsection{Positive-aware hard-negative mining methods}
\label{sec:mining_methods} 

A popular contrastive learning loss for retrieval models is \textit{InfoNCE} \cite{oord2018representation} shown in Equation~\ref{eq:infonce}, where $ sim(\cdot) $ is a similarity function (e.g., cosine similarity or dot product), $ d^+ $ is a positive relevant passage, $ d^-$ is one of the $ N $ negative passages and $ \tau $ is the temperature parameter. The objective is to maximize the embedding similarity between the query and positive passage, whilst minimizing the similarity between the query and negative passages.

\begin{equation*}
\mathcal{L}(q, d^+, d_N) = -\log p(d = d^+ \mid q) 
\end{equation*}   
\begin{equation}    
    = -\log \frac{\exp(\text{sim}(q, d^+)/\tau)}{\sum_{d_i \in \{d^+\} \cup d_N} \exp(\text{sim}(q, d_i)/\tau)},
\label{eq:infonce}
\end{equation}   

As discussed in Section~\ref{sec:hard_neg_mining_background}, random negatives are uninformative for training and hard-negatives (with higher relevance score $ sim(q,d_i) $) can provide faster convergence. 

The basic method for mining hard-negatives is to select the top-k most similar candidates to the query (ignoring positive passages) which we name \textbf{\textit{Naive Top-K}}.

However, the hard-negative mining process may introduce some false negatives, i.e., passages $d_i$ are relevant to the query but not annotated as positives $d^+$, which add noise to constrastive learning. 
False negatives can be present after negative mining when annotation of positive passages is not comprehensive enough or when mining is done on a large corpus. For example, in open-domain question answering (OpenQA) datasets such as MS MARCO \cite{bajaj2016ms} and Natural Questions \cite{kwiatkowski2019natural}, the question answer might be supported by many paragraphs from Wikipedia\cite{chen2017reading,kwiatkowski2019natural} or the web \cite{bajaj2016ms}. 

Some methods for filtering out false negatives have been proposed in the literature:

\begin{itemize}    
    \item \textbf{\textit{Top-K shifted by N}} - Selects the top-k negatives after a rank $N$, e.g., Top-10 shifted by 5 would ignore the first 5 negatives and consider negatives between rank 5 and 15 \cite{bge_embedding, SFRAIResearch2024};
    \item \textbf{Top-k with absolute threshold (\textit{TopK-Abs})} - Ignores negatives with relevance score higher than an absolute threshold \cite{qu2020rocketqa, merrick2024arctic, lee2024nv};
\end{itemize}

These methods have some important limitations.  \textit{Top-K shifted by N} is a basic method that does not take into account the relevance score of the negative with respect to the query and might even throw away valuable hard-negatives or keep false negatives. \textit{TopK-Abs} uses absolute thresholds as maximum negative scores with respect to the query, regardless the positive passage relevance.

Motivated by those limitations, we have designed a family of \textit{positive-aware hard-negative mining methods}. Our methods are simple, generic and can be applied on top of any teacher model (e.g., embedding, reranker) that retrieves top-k candidates semantically relevant to the query. Our methods take advantage of the information from the positive relevance score to help identify and eliminate potential false negatives. The base method is described in Algorithm~\ref{alg:pos_mining_method}. It iterates over the retrieved top-k negatives and filters out potential false negatives using one of the following filter criteria:

\begin{itemize}   
    \item \textbf{Top-k with margin to positive threshold (\textit{TopK-MarginPos})} - The maximum threshold for negative scores is the positive score minus an absolute margin (Algorithm~\ref{alg:TopKMarginPos}).
    \item \textbf{Top-k with percentage to positive threshold (\textit{TopK-PercPos})} - The maximum threshold for negative scores is a percentage of the positive score (Algorithm~\ref{alg:TopKPercPos}).
\end{itemize}

\begin{algorithm}
\caption{Positive-aware hard-negative mining base method}\label{alg:pos_mining_method}
\begin{algorithmic}[1]
\Procedure{PositiveAwareNegativeMining}{$p,N$}

\State \text{valid\_negatives} = []

\For {all $n$ in negatives $N$}
    \If{\textit{filter\_fn}($p$,$n$)} \Comment{Call one of the positive-aware mining filters: \textit{TopKMarginPosFilter} or \textit{TopKPercPosFilter}}
        \State valid\_negatives.append(n) 
    \EndIf
\EndFor
\State \textbf{return} \text{valid\_negatives}
\EndProcedure
\end{algorithmic}
\end{algorithm}

\begin{algorithm}
\caption{TopK-MarginPos negatives filter}\label{alg:TopKMarginPos}
\begin{algorithmic}[1]
\Procedure{TopKMarginPosFilter}{$p,n$}
\State abs\_margin $\gets$ <CONFIG\_MARGIN>
\State \textbf{return}$\gets$ ($n$.rel\_score <
$p$.rel\_score - abs\_margin)
\EndProcedure
\end{algorithmic}
\end{algorithm}

\begin{algorithm}
\caption{TopK-PercPos negatives filter}\label{alg:TopKPercPos}
\begin{algorithmic}[1]
\Procedure{TopKPercPosFilter}{$p,n$}
\State perc\_margin $\gets$ <CONFIG\_MARGIN>
\State \textbf{return} $\gets$ ($n$.rel\_score <
$p$.rel\_score * perc\_margin)
\EndProcedure
\end{algorithmic}
\end{algorithm}

In Section~\ref{sec:experiments}, we compare mining methods by fine-tuning different base models on negatives mined using distinct teacher models, and demonstrate the effectiveness of our proposed \textit{positive-aware mining methods} at scale to achieve state-of-the-art retrieval accuracy. We also provide an ablation study on the margin configurations of our methods to serve as a reference.

\subsection{Research Questions}
\label{sec:methodology}



In this paper, we investigate the following Research Questions with respect to hard-negative mining:

\begin{itemize}    
    \item \textit{RQ1}. How much does mining hard-negatives with different teacher models affect the accuracy of the downstream fine-tuned embedding models?
    \item \textit{RQ2}. Can ensembling hard-negatives mined from different teacher models improve results? 
    \item \textit{RQ3}. How does different hard-negative mining methods for fine-tuning compare on the evaluation accuracy?   
\end{itemize}

The next sections present the experiments results for those research questions, comparing different teacher models for mining and the effect of ensembling hard-negatives from distinct models. We also provide a comprehensive experimentation on hard-negative mining methods over their thresholds for filtering out false negatives.





\subsection{Experiments setup}

\subsubsection{Training} 

In our general setup, embedding models are finetuned\footnote{We use for fine-tuning https://github.com/microsoft/unilm/tree/master/simlm} with constrastive learning  from \textit{e5-large-unsupervised} or \textit{Mistral-7B-v0.1} base models, with hard-negatives mined using selected mining methods and teacher models of different sizes.


Train set is composed of Natural Questions (NQ) \cite{kwiatkowski2019natural} \footnote{https://ai.google.com/research/NaturalQuestions}, Stack Exchange (2023
 dump) \footnote{https://archive.org/details/stack-exchange-data-dump-2023-09-12} and SQUAD \cite{rajpurkar2018know} \footnote{https://rajpurkar.github.io/SQuAD-explorer/} datasets (~287k examples).

\subsubsection{Evaluation} 
We selected three Question-Answering datasets from MTEB Retrieval / BEIR benchmark - NQ, HotpotQA and FiQA-2018 \cite{thakur2021beir} - as they are more relevant for Q\&A RAG systems. 

For \textit{RQ3}, we also perform a scaled experiment setup using a larger train set (same from \textit{NV-Retriever-v1} model) and evaluate on full MTEB Retrieval benchmark, as described in Section~\ref{sec:nv_retriever_ablation}.







\section{Experiment results and discussion}
\label{sec:experiments}

Here we investigate the research questions presented in Section~\ref{sec:methodology}.

\subsection{RQ1. Using different teacher models for mining}
\label{sec:exp_models}

We have selected  a number of popular text embedding models as teacher models for mining hard-negatives. Those models represent different architectures, model sizes and retrieval accuracy.


\begin{itemize}    
    \item \textbf{\textit{e5-large-unsupervised}} \footnote{https://huggingface.co/intfloat/e5-large-unsupervised} (334M params) - The \textit{E5} model pre-trained on unsupervised data with CL \cite{wang2022text};
    \item \textbf{\textit{e5-large-v2}} \footnote{https://huggingface.co/intfloat/e5-large-v2} (334M params) - An \textit{E5} model fine-tuned on top of \textit{e5-large-unsupervised} with supervised data\cite{wang2022text} 
    \item \textbf{\textit{snowflake-arctic-embed-l}} \footnote{https://huggingface.co/Snowflake/snowflake-arctic-embed-l} (334M params) - Member of the \textit{artic-embed} models which is trained in two rounds (with supervised and unsupervised data) like \textit{E5} model, with improvements on data and training that lead to higher retrieval accuracy \cite{merrick2024arctic} 
    \item \textbf{\textit{e5-mistral-7b-instruct}} \footnote{https://huggingface.co/intfloat/e5-mistral-7b-instruct} (7.1B params) - The decoder-only Mistral model fine-tuned with CL to create an embedding model \cite{wang2023improving}; 
    \item \textbf{\textit{NV-Embed-v1}} \footnote{https://huggingface.co/nvidia/NV-Embed-v1} (7.8B params) - A Mistral-based embedding model with some modifications including bi-directional and latent attention\cite{lee2024nv}. 
\end{itemize}

We mine 4 hard-negatives\footnote{For the experiments comparing different mining models, in order to remove potential false negatives, we use the \textit{TopK-PercPos} mining method configured to set the maximum threshold for hard-negative scores as 95\% of the corresponding positive score. That mining method and configuration performed best in our ablation, as described in Section~\ref{sec:comparing_methods}.} with those different teacher models for every question in the train set. This process results in one train set per teacher model, which is then used for fine-tuning the base model (\textit{E5-large-unsupervised}).

We can see from the results on Table~\ref{tab:ablation_diff_embeddings} that the worse retrieval accuracy was obtained by using negatives mined with the BM25 sparse retrieval model\cite{robertson2009probabilistic} followed by random negatives, which was surprising and opposite to what was found in \cite{karpukhin2020dense}.
The \textit{E5-large-unsupervised} dense retrieval model, pre-trained only on unsupervised data from the web, provided better accuracy. 

The next teacher models are trained on retrieval supervised data, particularly for Question-Answering RAG systems. The \textit{e5-large-v2} and \textit{snowflake-arctic-embed-l} use the \textit{E5} architecture (334M params) and perform better than the baselines. 
The best teacher models were the larger \textit{NV-Embed-v1} and \textit{e5-mistral-7b-instruct}, both based on the Mistral 7B architecture. They provide better hard-negatives for CL, which result on higher accuracy for the fine-tuned models.

\begin{table}[ht]
\caption{Evaluation (NDCG@10) of fine-tuned \textit{e5-large-unsupervised} embedding models with hard-negatives mined using different teacher models}
\footnotesize
\begin{tabular}{l|c|c|c|c}
Teacher models                                       & Avg.   & NQ     & HotpotQA & FiQA   \\ \hline \hline
BM25                         & \textbf{0.5002} & 0.5307 & 0.5774   & 0.3923 \\ 
random                                        & \textbf{0.5248} & 0.5123 & 0.6151   & 0.4471 \\ 
e5-large-unsupervised                         & \textbf{0.5494} & 0.5541 & 0.6247   & 0.4694 \\ \hline
e5-large-v2                                   & \textbf{0.5704} & 0.6058 & 0.6435   & 0.4618 \\
snowflake-arctic-embed-l                      & \textbf{0.5728} & 0.6118 & 0.6331   & 0.4735 \\
NV-Embed-v1                                   & \textbf{0.5744} & 0.6092 & 0.6355   & 0.4785 \\
e5-mistral-7b-instruct                        & \textbf{0.5810} & 0.6241 & 0.6434   & 0.4757 \\ 

\hline
\end{tabular}
\label{tab:ablation_diff_embeddings}
\end{table}

\subsection{RQ2. Ensembling hard-negatives from different teacher models}
\label{sec:ensembling_models}
Ensembling outputs from different models is a common practice in machine learning to improve predictions accuracy, for providing a more robust estimator \cite{deotte2021gpu,schifferer2021using,moreira2021transformers}.

In particular, we investigated the similarity of the top-4 hard-negatives  mined by four teacher models and noticed a low level of agreement (\textit{jaccard similarity} lower than 30\%), as you can see on Appendix~\ref{sec:ablation_diversity_negatives_models}. 
For this reason, we decided to explore ensembling to try improving the quality of the hard-negatives.

We explored two methods to combine hard-negatives mined from four different \textit{E5} and \textit{Mistral} based embedding models - \textit{e5-large-v2}, \textit{snowflake-arctic-embed-l}, \textit{NV-Embed-v1}, and \textit{e5-mistral-7b-instruct} - which are described next. Each ensembling method returns 4 hard-negatives for each example (query,positive).

\begin{itemize}
    \item \textit{Cross-sample ensembling} - For each example, samples a teacher model to obtain all the negatives.
    \item \textit{Intra-sample ensembling} - For each example, selects the top-1 mined negative from each teacher model.
\end{itemize}

The evaluation results are shown in Table~\ref{tab:ablation_ensemble}. In the first row, we have as baseline the model trained with hard-negatives from the best teacher model (from \textit{RQ1}): \textit{e5-mistral-7b-instruct}. 
We can see that the \textit{cross-sample ensembling} method does not lead to better negatives than the best teacher model.

The \textit{Intra-sample ensembling} method turned out to be more effective. It may lead to duplicate hard-negatives, as for some examples the teacher models might agree on the 1st hard-negative. Thus, 
we tried two variations: keeping duplicates (\textit{no-dedup}) or removing duplicates (\textit{dedup}) and replacing them by the next unique hard-negative from teacher models sorted by their accuracy. Surprisingly, we found that it was better to keep the duplicate hard-negatives for training. A possible explanation could be that if models agree on the 1st hard-negative, keeping it duplicate will increase its importance in the cross-entropy loss.

\begin{table}[ht]
\caption{Evaluation (NDCG@10) of fine-tuned \textit{e5-large-unsupervised} embedding model with ensembles of hard-negatives mined using 4 different models: \textit{e5-large-v2}, \textit{snowflake-arctic-embed-l}, \textit{NV-Embed-v1}, \textit{e5-mistral-7b-instruct}}
\footnotesize
\begin{tabular}{l|c|c|c|c}
Teacher Model / Ensemble method                                       & Avg.   & NQ     & HotpotQA & FiQA   \\ \hline \hline

e5-mistral-7b-instruct (baseline)                        & \textbf{0.5810} & 0.6241 & 0.6434   & 0.4757 \\ \hline
\multicolumn{5}{l}{\textit{Ensembled hard-negatives from 4 models}} \\ \hline
Cross-sample ensembling & \textbf{0.5806}  & 0.6279 &  0.6384  & 0.4611 \\ 
Intra-sample ensembling (dedup) & \textbf{0.5804} & 0.6324 & 0.6302   & 0.4716 \\ 
Intra-sample ensembling (no-dedup) & \textbf{0.5825} & 0.6357 & 0.6298   & 0.4820 \\ 

\hline
\end{tabular}
\label{tab:ablation_ensemble}
\end{table}

\subsection{RQ3.a Comparing methods for mining hard-negatives}
\label{sec:comparing_methods}

To investigate RQ3 we performed a comprehensive number of experiments testing the different hard-negative mining methods described in Section~\ref{sec:mining_methods}. 


\subsubsection{Ablation study on mining methods configurations}

We performed an ablation study where we fine-tune the \textit{e5-large-unsupervised} model (\textit{Base setup}) with hard-negatives mined by different mining methods, for a range of their configuration.  

For \textit{TopK-Abs}, \textit{TopK-MarginPos} and \textit{TopK-PercPos} methods, the interval for the threshold/margin values config was $[0,1]$, with increments of $0.05$\footnote{For \textit{TopK-PercPos}, we included $0.98$ and  $1.05$ thresholds for more fine-grained analysis around the best threshold.}. We use as teacher model the \textit{e5-mistral-7b-instruct}, which performed best for mining hard-negatives (Section~\ref{sec:exp_models}).


We present plots with the results for the different configuration choices of each mining method. The evaluation metric reported is the average of NDCG@10 for the three selected BEIR Q\&A datasets (NQ, HotpotQA and FiQA). 

\begin{figure}[ht]
    \centering
    \includegraphics[width=1.0\linewidth]{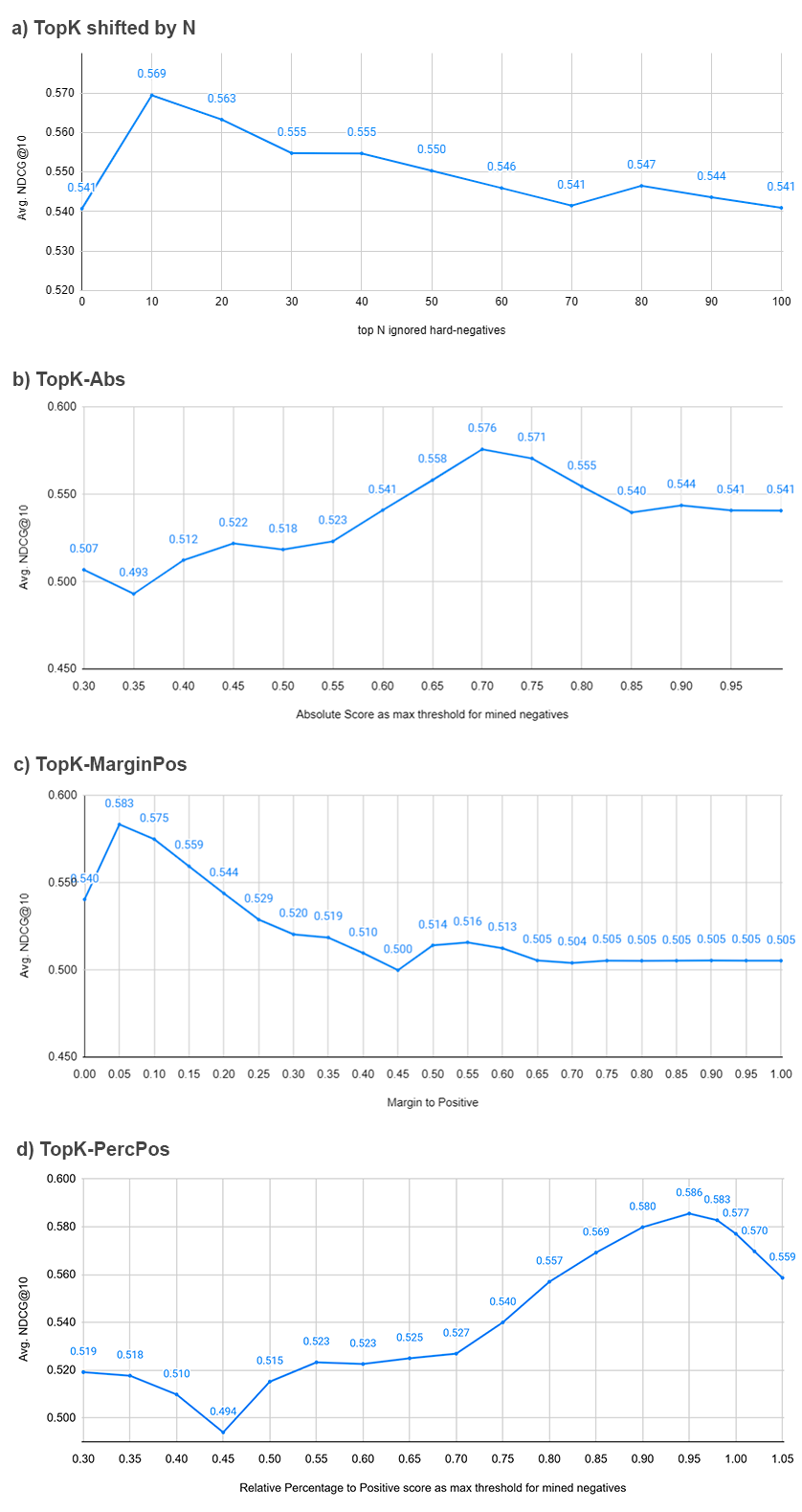}
    \caption{Ablation study of the negative mining methods over their configuration.}
    \label{fig:topk_ablation}
\end{figure}

We can see that the basic \textit{Top-k shifted by N} (Figure~\ref{fig:topk_ablation}.a) method provides its best accuracy when discarding the top-10 ranked hard-negatives from the training, as they may be too strong and have a higher chance of containing false negatives.

For \textit{TopK-Abs} (Figure~\ref{fig:topk_ablation}.b) -- a traditional method to remove mined false negatives -- the best configuration was to use the absolute score of 0.7 as the maximum threshold for negative scores. It seems that a higher threshold could include more false negatives, and a lower threshold could mine weaker negatives that would not be very informative for CL.

We propose in this paper \textit{positive-aware mining methods}: \textit{TopK-MarginPos} and \textit{TopK-PercPos}. For \textit{TopK-MarginPos} mining method (Figure~\ref{fig:topk_ablation}.c), we observe that subtracting a small margin (0.05) from the positive score and setting it as the maximum threshold for the hard-negatives helped remove potential false negatives, and larger margins end up penalizing the model accuracy. 
For the \textit{TopK-PercPos} mining method (Figure~\ref{fig:topk_ablation}.d), we find that setting the maximum threshold for mined negatives as 95\% of the positive scores was the optimum configuration. We can see that training on high-scoring negatives with respect to positives is detrimental to model accuracy, as it decreases after relative margin goes higher than 95\% threshold.

\subsubsection{Sampling from mined negatives}

After a mining method returns an ordered list of negatives, the typical approach is selecting the top-k candidates, but some research works have proposed sampling among top-k to add some relevance diversity among the selected hard-negatives:

\begin{itemize}        
    \item \textbf{\textit{Sampled Top-k}} - Samples $n$ negatives from the top-k most relevant ones \cite{qu2020rocketqa, lee2024gecko, chen2024bge} or from a range of negatives based on its relevance rank, e.g. from 30-100 range like in \cite{SFRAIResearch2024};
    \item \textbf{\textit{Top-1+sampled top-k}} - Selects the top-1 hard-negative (to secure a strong one) and samples $n-1$ negatives like described in the \textit{Sampled Top-k} method.
\end{itemize}

\begin{figure}[ht]
    \centering
    \includegraphics[width=1.0\linewidth]{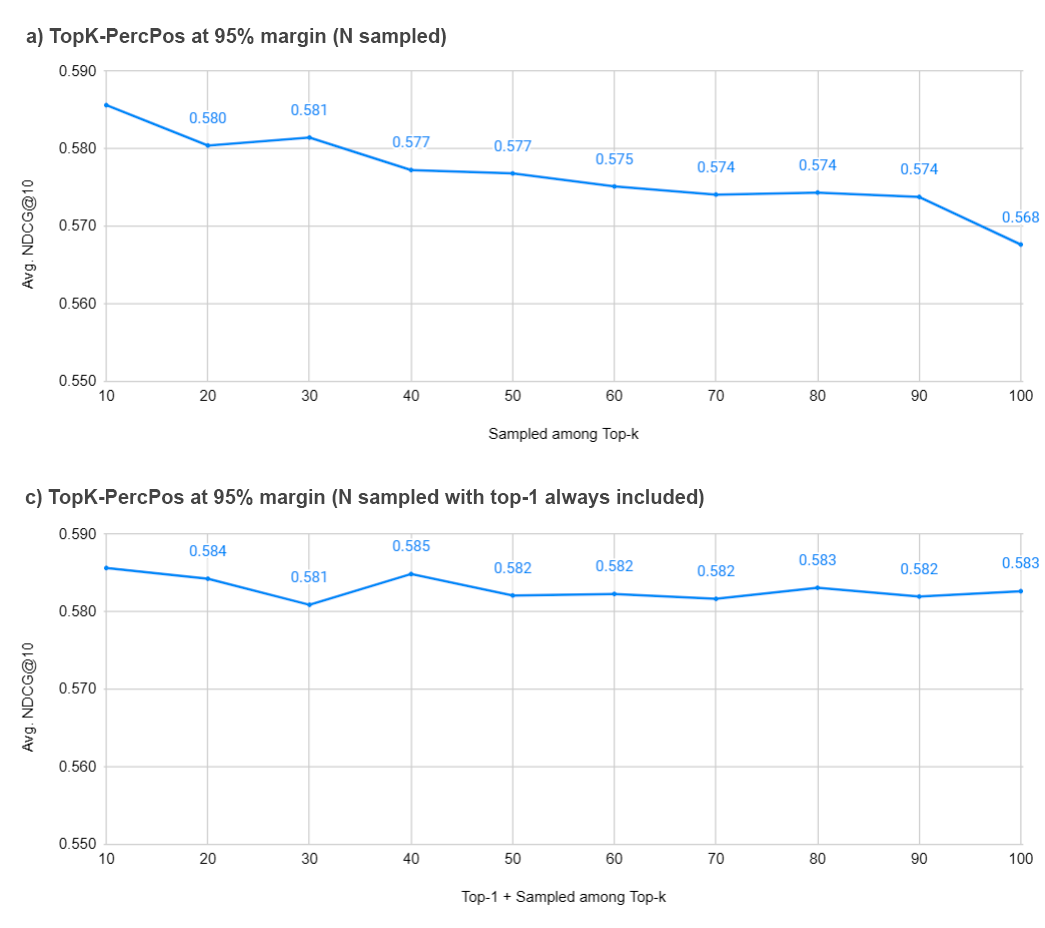}
    \caption{Ablation study of \textit{TopK-PercPos} negative mining method with 95\% threshold + (\textit{Sampled Top-k}) or (\textit{top1 + Sampled}) sampling method. Four negatives are sampled among different top $k$}
    \label{fig:topk_perc_pos_sampled}
\end{figure}

We performed experiments of those two sampling approaches using \textit{TopK-PercPos} (at 95\% config)\footnote{As we observed empirically that it is important to have strong negatives, in our sampling implementation we use \textit{softmax} to create a probability distribution based on the negatives relevance score and sample based on that distribution.}. We experimented with $k$ in the range of [10,100] with increments of 10, as shown in Figure~\ref{fig:topk_perc_pos_sampled}.

The best configuration was sampling the four negatives from top-10. Increasing the $k > 10$ for sampling dropped the retrieval accuracy for \textit{Sampled Top-k} (Figure~\ref{fig:topk_perc_pos_sampled}.a), but not much for \textit{Top-1+sampled top-k} sampling method (Figure~\ref{fig:topk_perc_pos_sampled}.b), as the latter ensures the top-1 mined negative is always included. 

\subsubsection{Comparing mining methods at their best configuration from the ablation}

We summarize in Table~\ref{tab:ablation_methods} the results of the best configurations of each mining method found in the ablation (Section~ \ref{sec:comparing_methods}). 

\begin{table}[ht]
\caption{NDCG@10 of fine-tuned \textit{e5-large-unsupervised} model with top-4 hard-negatives from different mining methods, at their best configuration found in the ablation.}
\footnotesize
\begin{tabular}{p{2.8cm}ccccc}
Mining Method                               & Config.     & Avg.   & NQ     & HotpotQA & FIQA   \\ \hline \hline
\textit{Naive Top-K}                                &      -       & 0.5407 & 0.5445 & 0.6120   & 0.4658 \\
\textit{Top-K shifted by N}                                &      N=10       & 0.5695 & 0.6007 & 0.6384   & 0.4693 \\
\textit{TopK-Abs}               & 0.7        & 0.5759 & 0.6133 & 0.6396   & 0.4748 \\
\textit{TopK-MarginPos}     & 0.05 & 0.5835 & 0.6338 & 0.6400   & 0.4766 \\
\textit{TopK-PercPos} & 95\%        & \textbf{0.5856} & \textbf{0.6369} & \textbf{0.6414}   & \textbf{0.4784} \\ \hline \multicolumn{6}{l}{\textit{Sampling method on negatives from \textit{TopK-PercPos} (at 95\% threshold)}} \\ \hline
\textit{TopK-PercPos (sampled)}                              & top-10      & 0.5856 & 0.6369 & 0.6414   & 0.4786 \\
\textit{TopK-PercPos (top1+sampled)}                        & top-10      & 0.5857 & 0.6369 & 0.6414   & 0.4787 \\ \hline
\end{tabular}
\label{tab:ablation_methods}
\end{table}

We can notice from the lower metrics of \textit{Naive Top-K} compared to the other mining models how important it is to remove the potential false negatives (noise) for fine-tuning. The basic \textit{Top-K shifted by N} method improves the obtained retrieval accuracy by ignoring the top (10) mined negatives, followed by \textit{TopK-Abs}, that ignores any negative scoring higher than a max. threshold (0.7).

Our \textit{positive-aware hard-negative mining methods} -- \textit{TopK-MarginPos} and \textit{TopK-PercPos} perform the best. In particular, the \textit{TopK-PercPos} was the most effective hard-negative mining method, with its threshold set to $95\%$ of the positive score. 

Sampling among top-k on top of \textit{TopK-PercPos} mining method did not help to further improve results for \textit{e5-large-unsupervised} base model, but it might help in some cases, e.g. for  Mistral 7B v0.1 base model, as we show in Table~\ref{tab:ablation_methods_mistral} and discuss in the next section.

\subsubsection{Replicating mining methods experiments with \textit{Mistral-7B-v0.1}}

In order to verify if our comparison of mining methods would generalize for base models other than \textit{e5-large-unsupervised} (334M params), we ran experiments fine-tuning another base model: \textit{Mistral-7B-v0.1} \footnote{https://huggingface.co/mistralai/Mistral-7B-v0.1} (7.1B params). As Mistral is a much larger model, we had to use only 1 hard-negative per example to avoid out of memory errors, and re-used the best configuration for each of the mining methods from the ablation with \textit{e5-large-unsupervised} (as running the full ablation with Mistral would require a lot of compute).

The mining methods comparison for \textit{Mistral-7B-v0.1} base model can be seen in Table~\ref{tab:ablation_methods_mistral}. It is noticeable the retrieval accuracy improvement when using Mistral larger base model compared to \textit{e5-large-unsupervised} (Table~\ref{tab:ablation_methods}) for fine-tuning. For \textit{Mistral-7B-v0.1} base model, we can see that \textit{Top-K shifted by N} also improves upon the \textit{Naive Top-K}, while the traditional \textit{TopK-Abs} mining method performs worse. On the other hand our \textit{positive-aware mining methods} perform much better, with \textit{TopK-PercPos} method also providing the best hard-negatives for finetuning.

\begin{table}[h]
\caption{NDCG@10 of \textit{Mistral-7B-v0.1} base model with top-1 hard-negatives from different mining methods, at the best configuration found in the ablation}.
\footnotesize
\begin{tabular}{llcccc}
Mining Method                               & Config.     & Avg.   & NQ     & HotpotQA & FIQA   \\ \hline \hline
\textit{Naive Top-K}                                &      -       & 0.6214 & 0.6450 & 0.6733   & 0.5458 \\
\textit{Top-K shifted by N}                                &      N=10       & 0.6342 & 0.6247 & 0.7126   & 0.5653 \\
\textit{TopK-Abs}          &   0.7  & 0.6184        & 0.6416 & 0.6744 & 0.5391 \\
\textit{TopK-MarginPos}     & 0.05 margin & 0.6457 & 0.6694 & 0.7113   & 0.5565 \\
\textit{TopK-PercPos} & 95\%        & \textbf{0.6479} & \textbf{0.6766} & \textbf{0.7053}   & \textbf{0.5618} \\ \hline \hline
\multicolumn{6}{l}{\textit{Sampling method on negatives from \textit{TopK-PercPos} (at 95\% threshold)}} \\ \hline
\textit{TopK-PercPos (sampled)}                              & top-10      & 0.6499 & 0.6763 & 0.7063   & 0.5671 \\ \hline
\end{tabular}
\label{tab:ablation_methods_mistral}
\end{table}

The last row in Table~\ref{tab:ablation_methods_mistral} applies the \textit{Sampled Top-k}\footnote{We don't include the results for \textit{Top-1+sampled top-k} sampling method in Table~\ref{tab:ablation_methods_mistral} because they match the baseline \textit{TopK-PercPos} method, as for the experiments with Mistral base model we could only use 1 hard-negative due to memory limitations} sampling method on top of \textit{TopK-PercPos} mining method. We can see that for Mistral base model it was slightly better to sample hard-negatives among top-10 instead of always using the top-1 negative.

\subsection{RQ3.b Comparing mining methods at scale with \textit{NV-Retriever-v1}}
\label{sec:nv_retriever_ablation}

In this section, we present experiments for RQ3 performed at scale. We demonstrate that our proposed mining methods are the secret sauce to build the \textit{NV-Retriever-v1} model, 1st place on the MTEB leaderboard when published. We leverage for these experiments the same setup we used for fine-tuning \textit{NV-Retriever-v1}. It uses \textit{Mistral-7B-v0.1} as base model and \textit{E5-Mistral-7B} as teacher model for mining hard-negatives according to each mining method (at their best configuration from the ablation study). We trained models on the 15 retrieval datasets used to train the first stage of NV-Retriever-v1  (Appendix~\ref{sec:nvretriever_trainsets}) with a total of 728,160 examples. The model and training hyperparameters are the same from \textit{NV-Retriever-v1}, as detailed in Appendix~\ref{sec:nv_retrieval_v1}.

This set of experiments demanded a lot of compute resources for requiring (1) mining negatives with a large teacher model (\textit{E5-Mistral-7B}) for each mining method and train set, (2) fine-tuning large Mistral-7B models and (3) evaluating on full MTEB Retrieval.

We present in Table~\ref{tab:nvretriver_ablation} the results for the models trained with the different mining methods, evaluated on the 15 datasets of the MTEB Retrieval benchmark. We can see in this scaled setup an even larger retrieval accuracy advantage of our proposed positive-aware mining methods compared to the baseline methods. 

In particular, the model trained with \textit{TopK-PercPos} mining method achieves the highest Avg. NDCG@10 ($60.55$), which would place 1st on the MTEB Retrieval / BEIR leaderboard leaderboard when \textit{NV-Retriever-v1} model was published (see Appendix~\ref{sec:mteb_results}). The \textit{NV-Retriever-v1} model scored only slightly higher ($60.9$) on MTEB because after the first stage training on those retrieval datasets, it is further fine-tuned with some classification and clustering datasets\footnote{The reason for not including those non-retrieval train sets in the experiments is because hard-negative mining is performed only for retrieval datasets.}. Our \textit{TopK-PercPos} method provided the best retrieval accuracy for 13 of the 15 BEIR datasets, for the other 2 datasets (\textit{SCIDOCS} and \textit{Touche-2020}) our \textit{TopK-MarginPos} method was the best.

This set of large experiments makes it is evident how sensitive the accuracy is to the choice of a hard-negative mining method, and how our positive-aware methods are effective in helping to achieve state-of-the-art results in MTEB Retrieval benchmark, as demonstrated by \textit{NV-Retriever-v1}.

\begin{table*}[htb]
\caption{\textit{Scaled setup}: Evaluation (NDCG@10) on full MTEB Retrieval of models trained with the same setup as \textit{NV-Retriever-v1} using  different hard-negative mining techniques}
\footnotesize
\begin{tabular}{lp{0.3cm}p{0.7cm}p{0.7cm}p{0.7cm}p{0.7cm}p{0.7cm}p{0.7cm}p{0.7cm}p{0.7cm}p{0.7cm}p{0.3cm}p{0.7cm}p{0.7cm}p{0.7cm}p{0.7cm}p{0.7cm}}
Mining method         & Avg                                & ArguAna                           & Climate-FEVER                      & CQA-Dupstack-Retrieval              & DBPedia                           & FEVER                             & FiQA2018                          & Hotpot-QA                          & MS-MARCO                           & NF-Corpus                          & NQ                                & Quora-Retrieval                    & SCIDOCS                           & SciFact                           & Touche-2020                        & TREC-COVID                         \\ \hline \hline
Naive Top-K           & \multicolumn{1}{r}{51.44}          & \multicolumn{1}{r}{63.8}          & \multicolumn{1}{r}{38.0}          & \multicolumn{1}{r}{41.6}          & \multicolumn{1}{r}{42.6}          & \multicolumn{1}{r}{89.2}          & \multicolumn{1}{r}{48.5}          & \multicolumn{1}{r}{73.9}          & \multicolumn{1}{r}{38.2}          & \multicolumn{1}{r}{13.1}          & \multicolumn{1}{r}{69.8}          & \multicolumn{1}{r}{84.8}          & \multicolumn{1}{r}{16.5}          & \multicolumn{1}{r}{77.6}          & \multicolumn{1}{r}{19.8}          & \multicolumn{1}{r}{54.2}          \\
Top-K shifted by (10) & \multicolumn{1}{r}{54.66}          & \multicolumn{1}{r}{57.3}          & \multicolumn{1}{r}{36.3}          & \multicolumn{1}{r}{45.8}          & \multicolumn{1}{r}{48.7}          & \multicolumn{1}{r}{87.6}          & \multicolumn{1}{r}{54.1}          & \multicolumn{1}{r}{75.2}          & \multicolumn{1}{r}{40.7}          & \multicolumn{1}{r}{31.1}          & \multicolumn{1}{r}{64.7}          & \multicolumn{1}{r}{86.4}          & \multicolumn{1}{r}{20.5}          & \multicolumn{1}{r}{73.0}          & \multicolumn{1}{r}{20.4}          & \multicolumn{1}{r}{78.0}          \\
TopK-Abs (0.7)        & \multicolumn{1}{r}{55.81}          & \multicolumn{1}{r}{61.7}          & \multicolumn{1}{r}{39.4}          & \multicolumn{1}{r}{49.7}          & \multicolumn{1}{r}{48.8}          & \multicolumn{1}{r}{90.4}          & \multicolumn{1}{r}{58.1}          & \multicolumn{1}{r}{74.1}          & \multicolumn{1}{r}{41.1}          & \multicolumn{1}{r}{24.1}          & \multicolumn{1}{r}{68.4}          & \multicolumn{1}{r}{88.2}          & \multicolumn{1}{r}{21.5}          & \multicolumn{1}{r}{76.2}          & \multicolumn{1}{r}{23.8}          & \multicolumn{1}{r}{71.5}          \\
TopK-MarginPos (0.05) & \multicolumn{1}{r}{59.77}          & \multicolumn{1}{r}{61.7}          & \multicolumn{1}{r}{39.9}          & \multicolumn{1}{r}{48.7}          & \multicolumn{1}{r}{50.5}          & \multicolumn{1}{r}{92.6}          & \multicolumn{1}{r}{\textbf{61.5}} & \multicolumn{1}{r}{77.7}          & \multicolumn{1}{r}{44.4}          & \multicolumn{1}{r}{44.6}          & \multicolumn{1}{r}{71.3}          & \multicolumn{1}{r}{88.4}          & \multicolumn{1}{r}{\textbf{22.5}} & \multicolumn{1}{r}{79.2}          & \multicolumn{1}{r}{\textbf{28.6}} & \multicolumn{1}{r}{84.8}          \\
TopK-PercPos (95\%)   & \multicolumn{1}{r}{\textbf{60.55}} & \multicolumn{1}{r}{\textbf{67.8}} & \multicolumn{1}{r}{\textbf{41.8}} & \multicolumn{1}{r}{\textbf{49.3}} & \multicolumn{1}{r}{\textbf{50.6}} & \multicolumn{1}{r}{\textbf{93.2}} & \multicolumn{1}{r}{\textbf{61.5}} & \multicolumn{1}{r}{\textbf{79.0}} & \multicolumn{1}{r}{\textbf{44.9}} & \multicolumn{1}{r}{\textbf{44.8}} & \multicolumn{1}{r}{\textbf{72.0}} & \multicolumn{1}{r}{\textbf{88.8}} & \multicolumn{1}{r}{22.1}          & \multicolumn{1}{r}{\textbf{80.0}} & \multicolumn{1}{r}{26.1}          & \multicolumn{1}{r}{\textbf{86.5}} \\ \hline
\end{tabular}
\label{tab:nvretriver_ablation}
\end{table*}

\subsection{The effects of positive-aware neg. mining}
In this section, we visualize the effects of  positive-aware hard-negative mining on false negatives removal and on shifting scores and loss distributions for faster constrastive learning.

\subsubsection{False negatives removal}
\label{sec:false_negs_removal}

To provide some insight on how effective our methods are in removing false negatives, we used LLM-as-a-judge (\textit{Llama 3.1 70b instruct}\footnote{We obtained very similar context relevance classification results using the same prompt with Mixtral 8x 22b, which increases the confidence of our LLM-as-judge approach.}) with a prompt to classify, for a sample of the train set with mined negatives, whether a context is relevant to the question (false negative). 

\begin{figure}[ht]
    \centering
    \includegraphics[width=0.9\linewidth]{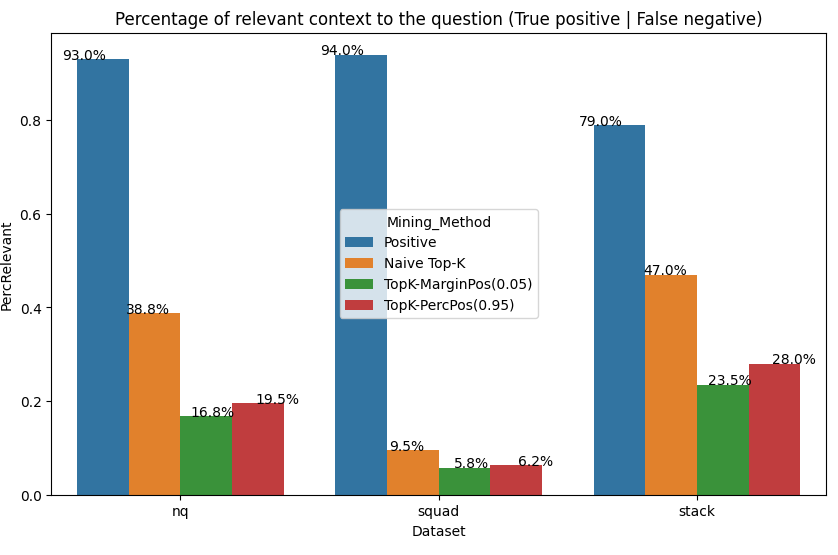}
    \caption{Percentage of relevant context - (true) positives and (false) mined negatives - as classified by LLM-as-a-judge (\textit{Llama 3.1 70b instruct}).}
    \label{fig:perc_relevant_context}
\end{figure}

We present in Figure~\ref{fig:perc_relevant_context} the percentage of relevant (false) negatives, together with the percentage of relevant (true) positives for reference. We observed that false negatives rate is generally proportional\footnote{Train set sizes: StackExchange (99,974), NQ (75,215), SQUAD (18,891)} to the number of unique passages in the corpus, as larger number of candidates increases the chance of relevant passages to be mined.
We can see that \textit{Naive Top-k} has a high false negative rate for NQ (38.8\%) and StackExchange (47\%) train sets as it selects as negatives the most similar contexts to the question without any filter. Our methods successfully mined 57\% and 50\% less false negatives than \textit{Naive Top-k} for NQ and StackExchange datasets, respectively. 

\subsubsection{Visualizing the effect of positive-aware mining}

\begin{figure*}[htb]
    \centering
    \includegraphics[width=0.90\linewidth]{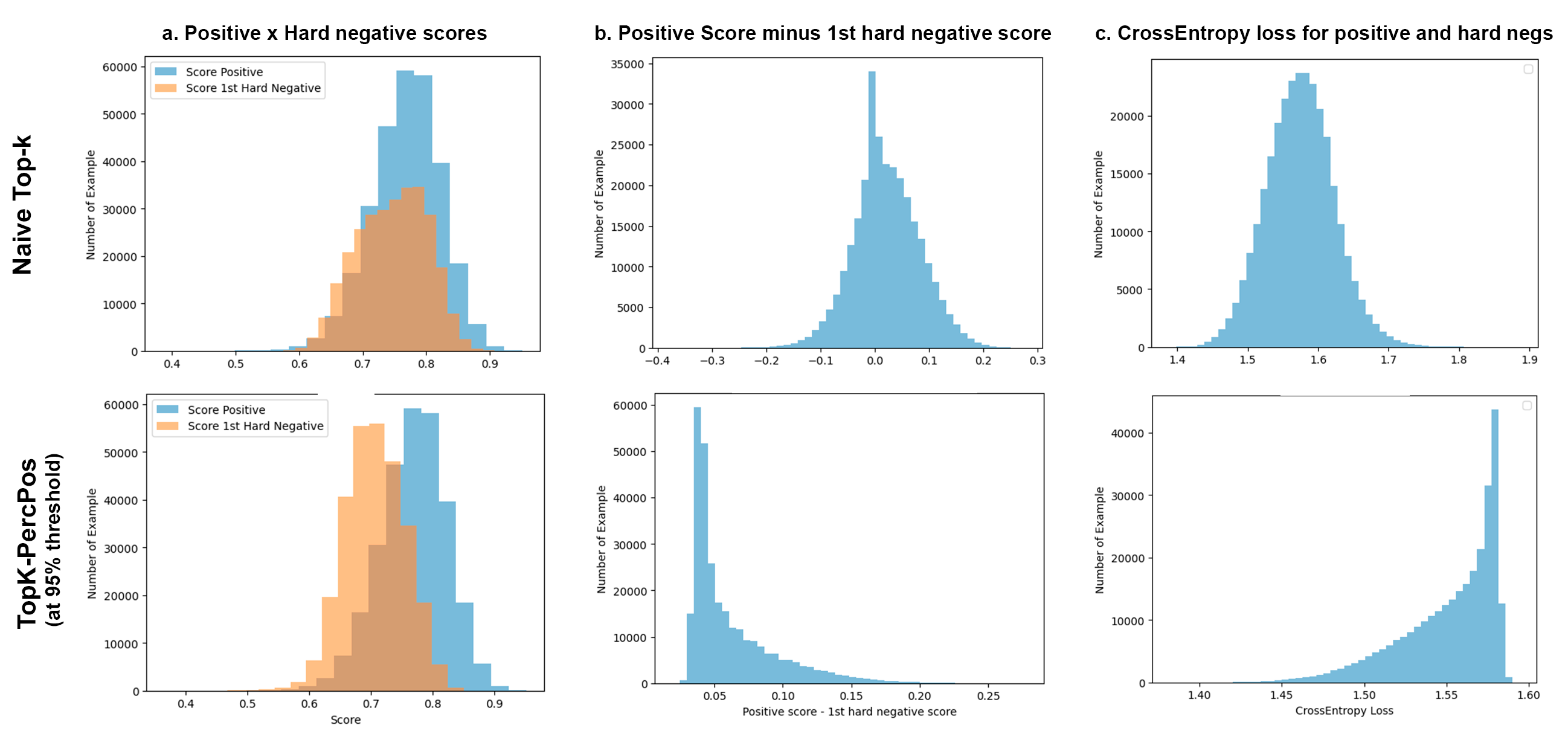}
    \caption{Histograms comparing \textit{Naive Top-k} and \textit{TopK-PercPos} mining methods}
    \label{fig:histograms_pos_negs}
\end{figure*}

To provide some intuition on the effect of removing potential false negatives, in Figure~\ref{fig:histograms_pos_negs} we provide histograms of the contexts scores and loss for \textit{Naive Top-k} and \textit{TopK-PercPos} mining methods on the train set. We can see in Figure~\ref{fig:histograms_pos_negs}.a that \textit{TopK-PercPos} helps to separate the distributions of positive and negative scores, in Figure~\ref{fig:histograms_pos_negs}.b that it avoids negatives scoring higher (negative difference) than the corresponding positive score, and in Figure~\ref{fig:histograms_pos_negs}.c that it limits the max. cross-entropy loss (Equation~\ref{eq:infonce}), making training more stable.

\section{Conclusion}

In this work, we introduce a novel family of \textit{positive-aware hard-negative mining methods}, that leverage the positive relevance score for mining better negatives for embedding models finetuning.

We provide a comprehensive ablation study comparing hard-negative mining methods (under their many configurations), different teacher models and the ensembling of their hard-negatives. We also investigate the effect of our methods on false negative removal, loss stabilization and accuracy improvement.
Finally, we demonstrate how effective our \textit{positive-aware hard-negative mining methods} are at scale, leading to the state-of-the-art \textit{NV-Retriever-v1} embedding model.

We recommend leveraging our mining methods for constrastive learning beyond text embedding models, as in some of our preliminary experiments on fine-tuning multi-modal (e.g., image,text) embedding models they also show to be effective in improving the retrieval accuracy. We suggest to future research works on text retrieval being aware of the sensitivity of models accuracy with respect to the negatives mined for contrastive learning. We also encourage them to
disclose their methodology for hard-negative mining for better reproducibility and replicability.

\bibliographystyle{ACM-Reference-Format}
\bibliography{manuscript}

\appendix

\section{Similarity of hard-negatives mined by different teacher models}
\label{sec:ablation_diversity_negatives_models}

In this appendix, we investigate the level of agreement of different teacher models on the hard-negatives.

In particular, we computed the Jaccard Similarity between the top-4 hard-negatives for pairs of different teacher models.  Table~\ref{tab:negatives_similarity} presents the similarity matrices for each train sets used in the ablation study: NQ, SQUAD and StackExchange. In general, it is possible to observe from the similarity matrices that the teacher models agree poorly on the top-4 mined hard-negatives, i.e., Jaccard similarity lower than 30\%.

\begin{table}[ht]
    \caption{Jaccard similarity of hard-negatives mined with different teacher models for NQ | SQUAD | StackExchange datasets}
    \footnotesize
    \begin{tabular}{p{1.5cm}|p{1.5cm}p{1.5cm}p{1.5cm}p{1.5cm}}
                             & e5-large-v2 & arctic-embed-l & NV-Embed-v1 & e5-mistral-7b \\ \hline

    e5-large-v2              & \multicolumn{1}{c}{-}           & 0.15 | 0.14 | 0.01                   & 0.11 | 0.13 | 0.01    & 0.11 | 0.14 | 0.01                 \\ \hline
        arctic-embed-l & 0.15 | 0.14 | 0.01    & \multicolumn{1}{c}{-}                        & 0.28 | 0.19 | 0.05    & 0.18 | 0.18 | 0.04                 \\ \hline
        NV-Embed-v1              & 0.11 | 0.13 | 0.01    & 0.28 | 0.19 | 0.05                   & \multicolumn{1}{c}{-}           & 0.23 | 0.30 | 0.09                 \\ \hline
        e5-mistral-7b   & 0.11 | 0.14 | 0.01    & 0.18 | 0.18 | 0.04                   & 0.23 | 0.30 | 0.09    & \multicolumn{1}{c}{-}                      \\ \hline

    \end{tabular}
    \label{tab:negatives_similarity}
    \end{table}

\section{NV-Retriever-v1}
\label{sec:nv_retrieval_v1}

In this section, we describe the architecture, methods and techniques for training the state-of-the-art \textit{NV-Retriever-v1} embedding model, which was placed first on the MTEB Retrieval leaderboard when it was published. The \textit{positive-aware mining methods} we propose in this paper were crucial to obtaining high retrieval accuracy for \textit{NV-Retriever-v1}, as we demonstrate in Section~\ref{sec:nv_retriever_ablation}.

\subsection{Model architecture}

The \textit{NV-Retriever-v1} uses the Mistral 7B\cite{jiang2023mistral}\footnote{https://huggingface.co/mistralai/Mistral-7B-v0.1 (we use the Mistral base version, without instruction fine-tuning)} as base model, as originally proposed by \textit{e5-mistral-7b-instruct} \cite{wang2023improving} and followed by many of the top performing embedding models on MTEB such as
\textit{Linq-Embed-Mistral}
\cite{LinqAIResearch2024}, \textit{GritLM} \cite{muennighoff2024generative},
\textit{SFR-Embedding-Mistral} \cite{SFRAIResearch2024}, and \textit{NV-Embed-v1} \cite{lee2024nv}. We convert the decoder \textit{Mistral-7b} model into an encoder model. We replace the causal attention of the base model with bi-directional attention, which empirically demonstrated higher accuracy in our experiment. Mean pooling is used to combine the outputs of the last Transformer layer, which corresponds to averaging the hidden states across the sequence length to form the sentence embedding. This approach was inspired by \textit{GritLM} \cite{muennighoff2024generative} and subsequent models using this approach like \textit{NV-Embed-v1} \cite{lee2024nv} and \textit{
gte-Qwen2-7B-instruct}.

\subsection{Train sets}
\label{sec:nvretriever_trainsets}

As MTEB contains different tasks like retrieval, reranking, classification, clustering among others, a diverse set of training datasets is required for a good overall performance. The train sets used for fine-tuning \textit{NV-Retriever-v1}, listed in Table~\ref{tab:datasets_training}, are based on those used to train \textit{E5-Mistral} \cite{wang2023improving} and \textit{NV-Embed-v1} \cite{lee2024nv}. We use the same instruction prompts from \cite{wang2023improving} for each dataset.

\subsection{Hard-negative mining}

We used the \textit{E5-Mistral-7B} embedding model for hard-negative mining\footnote{The only exception was ArguAna dataset, for which we found that using an internal Mistral embedding model for mining instead of \textit{E5-Mistral-7B} trained with causal attention improved significantly the NDCG@10 for Arguana test set.} with maximum sequence length of 4096. 

To ignore potential false negatives, we leverage our proposed \textit{TopK-PercPos} method and set the maximum threshold for the negative relevance score as 95\% of the positive score. We explain this choice in the ablation study presented in Section~\ref{sec:comparing_methods}.

To better leverage the base LLM model pre-training and adjust for datasets specific domain and task, \cite{wang2023improving} proposed and designed  specific natural language instructions for each train set. The instructions prefixes are added to the query but not to the passages, so that for the latter no re-indexing is required for different instructions. 
We also adopt instruction prefixes with a small difference in the implementation: instead of the original template from \cite{wang2023improving} "\textit{Instruct: \{task\_definition\} \textbackslash n Query: \{query\}}" , we use "\textit{\{task\_definition\}: \{query\}}". As in \textit{NV-Embed-v1} \cite{lee2024nv} we mask out the instruction tokens for the average pooling during both training and evaluation, which can still affect the output due to self-attention.

\begin{table}[htb]
\caption{Retrieval train datasets for \textit{NV-Retriever-v1}}
\footnotesize
\begin{tabular}{lr}
Dataset                                & \# of samples \\ \hline \hline
\textit{Retrieval datasets} & \\ \hline
ArguAna                                & 4065          \\
BioASQ                                 & 2495          \\
FEVER                                  & 50000         \\
FiQA2018                               & 14166         \\
GOOAQ                                  & 20000         \\
HotpotQA                               & 85000         \\
MS-MARCO                               & 200000        \\
NFCorpus                               & 3685          \\
Natural Language Inference             & 20000         \\
Natural Questions                      & 100231        \\
PAQ                                    & 20000         \\
SciFacts                               & 919           \\
SQUAD                                  & 87599         \\
StackExchange                          & 100000        \\
TriviaQA                               & 20000         \\ \hline
\textit{Non-retrieval datasets} & \\ \hline
Banking77Classification                & 10000         \\
AmazonCounterfactualClassification     & 4018          \\
AmazonReviewsClassification            & 20000         \\
EmotionClassification                  & 16000         \\
ImdbClassification                     & 15000         \\
MTOPIntentClassification               & 10000         \\
ToxicConversationsClassification       & 40000         \\
TweetSentimentExtractionClassification & 27481         \\
STS12                                  & 1868              \\
STS22                                  & 416              \\
STSBenchmark                           & 2812              \\
ArxivClusteringP2P                     & 35000         \\
ArxivClusteringS2S                     & 35000         \\
BiorxivClusteringP2P                   & 4070          \\
BiorxivClusteringS2S                   & 4070          \\
MedrxivClusteringP2P                   & 1160          \\
MedrxivClusteringS2S                   & 1160          \\ \hline
\end{tabular}
\label{tab:datasets_training}
\end{table}

\subsection{Training setup}
We perform two stages of instruction tuning as in \cite{lee2024nv}. In the first stage, we only use supervised retrieval datasets with in-batch negatives in addition to the mined hard-negatives. In the second stage, we blend that retrieval data with other tasks' datasets (e.g. classification, regression, semantic sentence similarity). 

Our training implementation is based on the Hugging Face Transformers\footnote{https://github.com/huggingface/transformers} and PEFT\footnote{https://github.com/huggingface/peft} libraries.  
The model and training hyperparameters are shown in Table~\ref{tab:model_hparams} and Table~\ref{tab:training_hparams}.

\begin{table}[ht]
\footnotesize
\caption{Model hyperparameters for \textit{NV-Retriever-v1}}
\begin{tabular}{lc}
\textbf{Hyperparameter} & \textbf{Value}            \\ \hline
Base model         & mistralai/Mistral-7B-v0.1 \\
Layers             & 32                        \\
Attention          & Bi-directional            \\
Embedding dim      & 4096                      \\
Embedding pooling  & Average at last layer     \\
LoRA Rank          & 16                        \\
LoRA Alpha         & 32                        \\
Query max length   & 192                       \\
Passage max length & 512                       \\ \hline
\end{tabular}
\label{tab:model_hparams}
\end{table}

\begin{table}[ht]
\caption{Training hyperparameters for \textit{NV-Retriever-v1}}
\footnotesize
\begin{tabular}{l|p{1.8cm}|l}
Hyperparameters             & 1st stage                           & 2nd stage      \\ \hline
Optimizer                   & \multicolumn{2}{c}{AdamW}                            \\
Learning rate               & \multicolumn{2}{c}{1.00E-05}                         \\
Learning rate warm-up steps & \multicolumn{2}{c}{100}                              \\
Negatives source            & hard-negatives + in-batch negatives & hard-negatives \\
Number of hard-negatives    & 1                                   & 5              \\
Batch-size                  & 32                                  & 8              \\
Gradient accumulation steps & 4                                   & 16             \\
Training Steps per epoch    & 2844                                & 829            \\
Epochs                      & 12                                  & 12             \\ \hline
\end{tabular}
\label{tab:training_hparams}
\end{table}

\subsection{Computational cost}
\label{sec:compute_cost}

NV-Retriever-v1 is trained in two stages, with 12 epochs each. Full training takes about 90 hours on 8x A100 GPUs. The hard-negative mining process depends on the corpus size and teacher model complexity. As a reference, for the \textit{e5-mistral-instruct} teacher model, it requires approximately 1.5 hours with 8x A100 GPUs to embed 1 million documents with max token length set to 4096. 

Evaluating an embedding model on the 15 MTEB BEIR retrieval datasets takes approximately 120 hours of 8x A100 GPUs.

\subsection{\textit{NV-Retriever-v1} results on MTEB Retrieval}
\label{sec:mteb_results}

\textit{NV-Retriever-v1} model achieves at MTEB leaderboard an average NDCG@10 score of $60.9$ and placed 1st place when published in July 11, 2024. Table~\ref{tab:mteb_results} reports the top embedding models on the MTEB Retrieval leaderboard when it was published. Mistral 7B was the foundation model for 4 of the top 5 places, that used similar train sets. As discussed before, the main difference in \textit{NV-Retriever-v1} training approach was the usage of the \textit{positive-aware mining methods} introduced in this paper.

\begin{table}[htb]
\footnotesize
\caption{Top embedding models on MTEB Retrieval leaderboard as of 2024-07-11}
\begin{tabular}{lp{0.7cm}p{0.5cm}p{0.5cm}p{0.8cm}p{0.5cm}p{0.5cm}}

MTEB Retrieval Dataset & NV-Retriever-v1 & gte-Qwen2-7B-instruct & Linq-Embed-Mistral & SFR-Embedding-2\_R & NV-Embed-v1    & SFR-Embedding-Mistral \\ \hline
\textbf{Average}       & \textbf{60.90}  & \textbf{60.25}        & \textbf{60.19}     & \textbf{60.18}     & \textbf{59.36} & \textbf{59.00}        \\ \hline
ArguAna                & 68.28           & 64.27                 & 69.65              & 62.34              & 68.20          & 67.17                 \\
ClimateFEVER           & 43.47           & 45.88                 & 39.11              & 34.43              & 34.72          & 36.41                 \\
CQADupstackRetrieval   & 49.36           & 46.43                 & 47.27              & 46.11              & 50.51          & 46.49                 \\
DBPedia                & 50.82           & 52.42                 & 51.32              & 51.21              & 48.29          & 49.06                 \\
FEVER                  & 93.15           & 95.11                 & 92.42              & 92.16              & 87.77          & 89.35                 \\
FiQA2018               & 61.18           & 62.03                 & 61.20              & 61.77              & 63.10          & 60.40                 \\
HotpotQA               & 79.12           & 73.08                 & 76.24              & 81.36              & 79.92          & 77.02                 \\
MSMARCO                & 44.89           & 45.98                 & 45.21              & 42.18              & 46.49          & 43.41                 \\
NFCorpus               & 45.06           & 40.60                 & 41.62              & 41.34              & 38.04          & 41.88                 \\
NQ                     & 72.44           & 67.00                 & 70.63              & 73.96              & 71.22          & 69.92                 \\
QuoraRetrieval         & 88.78           & 90.09                 & 90.27              & 89.58              & 89.21          & 89.78                 \\
SCIDOCS                & 22.55           & 28.91                 & 21.93              & 24.87              & 20.19          & 19.91                 \\
SciFact                & 81.31           & 79.06                 & 78.32              & 85.91              & 78.43          & 77.66                 \\
Touche2020             & 26.60           & 30.57                 & 30.61              & 28.18              & 28.38          & 29.00                 \\
TRECCOVID              & 86.44           & 82.26                 & 87.10              & 87.27              & 85.88          & 87.60                 \\ \hline
\end{tabular}
\label{tab:mteb_results}
\end{table}

\end{document}